\documentclass{aa}
\usepackage{psfig}

\usepackage{graphics}
\usepackage{psfig}
\begin{document}
\thesaurus{08(09.10.1;09.13.2;08.16.5)}

\title{Photometry of Nova  V 1493 Aql  
\thanks{Based on data collected at the Osservatorio
Astrofisico di Catania, stazione M. G. Fracastoro, Serra la Nave
(Etna), Italia}}
%\subtitle{}
\author{Piercarlo Bonifacio\inst{1}
\and Pierluigi Selvelli\inst{2}
\and Elisabetta Caffau\inst{3}}
\offprints{P. Bonifacio}
\institute{
Osservatorio Astronomico di Trieste,Via G.B.Tiepolo 11, 
I-34131 Trieste, Italia
\and
C.N.R.--G.N.A.--
Osservatorio Astronomico di Trieste,Via G.B.Tiepolo 11, 
I-34131 Trieste, Italia
\and
Istituto Magistrale S.P.P. e L. annesso al Convitto Nazionale 
Paolo Diacono, S. Pietro al Nat.
Udine, Italia
}
\mail{bonifaci@ts.astro.it}
\date{received .../Accepted...}
\maketitle

\begin{abstract}
  
We report on photometric observations
of V 1493 Aql during the early decline and highlight some 
uncommon aspects of the light curve . 
V 1493 Aql was hotter at maximum light than in 
the following phases, and was characterized by the presence 
of a long lasting secondary maximum, that, 
unlike in other novae,  was quite red in color.   
The mean of three distance estimates
yields  d $\sim 18.8\pm 3.6$ Kpc.
Such a large distance would 
place  V 1493 Aql at the extreme outskirts of our Galaxy or even 
in an external Local Group galaxy.

\end{abstract}

 
\section{Introduction}

V 1493 Aql (= Nova Aql 1999 n. 1)  was discovered   by
Tago (1999) as an 8.8 mag object  on two films
taken on July ~13.56 UT with a 55-mm f/3 camera lens.  
Nothing was visible on a film taken 4 days before. 
The precursor of the nova was too faint in quiescence 
to be recorded by the 1.2 m Palomar Schmidt, which sets the 
amplitude of the outburst in the $\delta {\rm m} \ge 12$ mag range (Moro 
et al, 1999).
Low-resolution spectra taken on July 14.6 UT 
by Ayani and Kawabata (1999) showed strong, broad Balmer lines 
with FWHM about 3400 ${\rm kms^{-1}}$. Similar values are reported by Tomov 
et al (1999) on July  ~15.9 UT who, in addition, remark the 
presence of FeII emission lines and the lack of absorption 
components in the Balmer emission lines. The decline of the nova 
was very fast and already  on July 17.02 UT  was fainter than 
11.0, thus indicating a  $t2$ of the order of 3 days.
 On August 3.89 the nova was already at 
$V=13.0$  (Lehky, 1999). 
We took advantage of an
observing run at the Catania Observatory to monitor the
following stages in the decline of the nova.
The object underwent a secondary outburst which began
shortly before our observations.
In this letter we  report our and other photometric
observations of V 1493 Aql  and discuss the implication of these
data in the  context of the classical nova phenomenon.

\section{Observations and data reduction}

\begin{figure*}
\psfig{figure=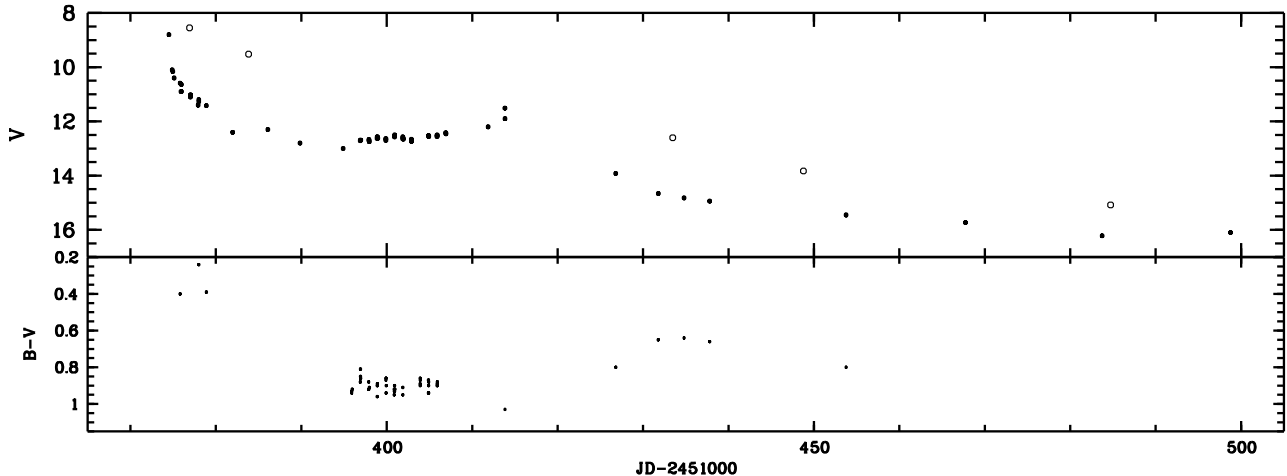,width=17.cm,angle=-90,clip=t}
\caption{Photometric data of V 1493 Aql as a function of time.
Open symbols are unfiltered CCD magnitudes, which usually 
calibrate onto Cousins R to within 0.1 mag. 
Filled symbols are either $V$ magnitudes 
(photoelectric or CCD) or photographic or visual estimates.
The data which is not from our observations is from
IAUC 7223 (Tago), IAUC 7225 (Hornoch, Pereira, Bouma,
Souza), IAUC 7228 ( Zejda, Safar, Masi, Hajek, Hanzl),
IAUC 7232 (Bouma, Reszelski, Schmeer, Hornoch, Lehky),
IAUC 7254 (Hanzl) IAUC 7273 (Yoshida \& Kadota, Hanzl, Masi),
IAUC 7313 (Zejda, Hanzl Masi) and one measure from
Masi (unpublished)}
\label{v}
\end{figure*}

From August 5th 1999 to August 15th
we observed V 1493 Aql with the 91 cm telescope of the 
Osservatorio Astrofisico di Catania at the 
M. G. Fracastoro mountain station on Mt. Etna, equipped
with a single
channel photometer with $UBV$ filters.
Standard stars from
the lists of Landolt (1983,1992) were observed each night
for calibration purposes.
The full data, as well as further
details on the observations,
are available in electronic form.  

Each night we observed at least two among four
stars which are angularly near to V 1493 Aql 
(GSC 01048-0098,
GSC 01048-01359, HD 230704, HD 178263) for comparison.
Unfortunately all the four  stars showed variations
in our photometry which
were above the expected measurement errors.
The Tycho Catalogue (Perryman et al 1997)
reports a photometric variability for all of them.
We made no use of the differential photometry, 
although all the data
is available electronically. 
Figure 1 shows  photometric data 
from the present work and from the IAU circulars,
while Figure 2 shows the Serra La Nave data in more detail.
In the time span by our observations  V 1493 Aql 
showed a brightening of about 0.4 mag.
Fluctuations from night to night and even within the night 
were larger than the estimated errors.
We searched for a periodicity in the photometric data
and may safely exclude the presence of periodicities
of 5 days, or shorter. 
The periodograms relative to $U$, $B$ and $V$
are very similar.

\section{The photometric behavior during decline}

It is well known that  each nova has its own
individuality and that no two novae are  completely alike in
their  photometric behavior (Payne--Gaposhkin
1957; Mc Laughlin 1960; Duerbeck 1981; Van den Bergh and
Younger 1987). Light curves 
tend to differ  from each other immediately after
the first phase of near exponential decay from maximum light
 showing either regular or irregular
variations  with oscillations, rapid decline (absorption)
followed by recovery, re-brightening, and in many  cases even a
continuous nearly exponential   smooth decline.  See also
Bianchini and Friedjung (1992) for an interpretation of the
oscillatory behaviour.

The light curve of V 1493 Aql is characterized, after the
first rapid decline to $V\sim   12$, by some evidence of the onset
 of small amplitude oscillations with the presence
of two local  minima at
$V =12.4$ (JD 381.904, Reszelski 1999) and at $V = 13.0$ 
(JD 394.89 Lehky, 1999).
Then a conspicuous re-brightening, with some indication of a
secondary oscillation takes place, almost in coincidence with
our observations.
The secondary maximum reaches $V =11.5$ on JD $\sim  420$  
(Hanzl, 1999).
AAVSO data  confirm the general  
trend near the secondary maximum, but do not show any clear 
evidence of oscillations in the previous phase.  
It is possible that the rather large scatter in these data has
masked  oscillations of small amplitude. 
If we consider the whole light curve,  
the brightening looks like a
secondary outburst with an increase in $V \sim 1.5$ mag  with
respect to the local minimum of $V=13.0$. 
From another perspective
the secondary maximum at $V = 11.5$
falls about 2.5 mag 
above the curve which corresponds to a 
smooth  decline. 
The total duration of the secondary outburst  is of
the order of 40 days.

\begin{figure}
\psfig{figure=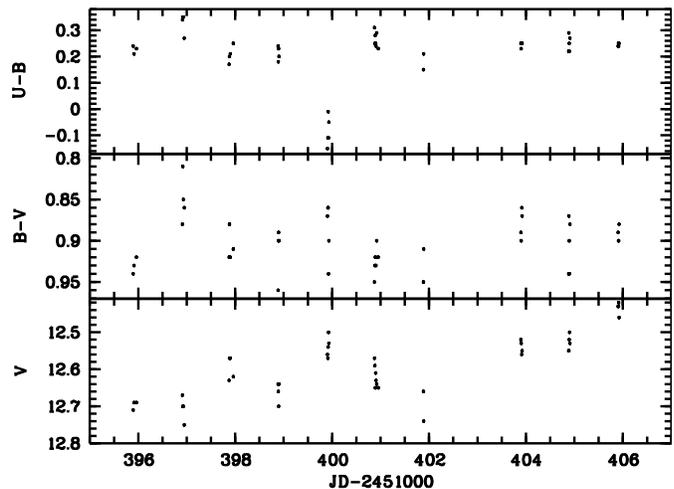,width=8.8cm,clip=t}
\caption{Serra La Nave data of V 1493 Aql}
\label{col}
\end{figure}

We have examined  several  light curves of novae in outburst, 
looking for similarities with 
V 1493 Aql and found only a marginal evidence for a similar
behaviour. While the presence of more or less coherent
oscillations is a characteristic shared by about 15 \% of novae
(with V 603 Aql and GK Per as prototypes), the long lasting
re-brightening ``profile'' of V 1493 Aql 
is quite rare, if not unique. To the best of
our knowledge, only a few  objects  bear some
resemblance   to     V 1493 Aql:
%\begin{enumerate}
% \item  
1) N Dor  1971a   (Van den Bergh \& Younger 1987), but the 
brightening was of much shorter duration and smaller  range in 
magnitude, and $t2$ much longer;
% \item 
2) N Cep 1971 (ibidem) , similar to N Dor 1971 but with some
evidence of an another small amplitude brightening;
%\item 
3) N Her 1963 = V533 Her, the same,  with additional small
amplitude brightenings;
%\item 
4) DK Lac (very long $t2$) presented many brightening
peaks of very short duration at semiregular time intervals;
%\item 
5) the recurrent nova T CrB showed a secondary maximum 
that took place about 100 days after the principal eruption.

%\end{enumerate}

Also the $(B-V)$ observations, despite
their coverage  of  the time interval of the nova
decline being only partial,  indicate a peculiar behaviour.
Other novae generally become bluer as they evolve :$(B-V)$ 
smoothly decreases  indicating that the shrinking in radius of 
the emitting  ``photosphere'' is accompanied by a gradual 
increase in temperature.  This is generally interpreted as a 
gradual decrease in the opacity, due to a decrease in density,
and the gradual uncovering of the hotter internal layers in the 
burning shell.  Instead, V 1493 Aql is hotter at maximum light ( 
$(B-V)<0.4$) than in the following phases: $(B-V)\sim  1$ near 
the secondary maximum  and $\sim  0.7$ in the following decline.
This clearly indicates that the secondary maximum corresponds to 
a relative increase of the size of the ``photosphere'' and a
decrease in temperature.
Only five points, quite well spaced  from the
beginning of the outburst up to   late stages,    are available for
the R band (Fig.  2), although,   none 
is near the secondary maximum.  
Remarkably,  they
define a curve which has a similar  slope as  the $V$ curve.
Actually, at a more accurate inspection (V-R) shows a decrease
with time, a behavior that contrasts with that generally
observed in  other 
novae, e.g. Nova Vul 1984 ( Robb and Scarfe , 1995), where 
$(V-R)$ steadily increases with the progress of the outburst.

\section{Absolute magnitude,  reddening and  distance}

Nova distances can be  derived  from a 
comparison between their intrinsic and observed  luminosities  
and therefore require that the latter are
corrected for the effect of interstellar extinction. 
Van den Bergh and Younger (1987) from an exhaustive 
$UBV$ study of novae at maximum, found that after
correction for reddening, the intrinsic color $(B-V)_0$  of novae
two magnitudes below maximum, i.e. at time $t2$, is
$(B-V)_0= -0.02 \pm 0.04$, with an rms  of 0.12.
Comparison with the measured $(B-V)$ allows to  estimate $E(B-V)$.
For V 1493 Aql we have two points near $t2$  with  $(B-V)=0.24$ and
$(B-V)=0.39$ respectively. If we assume  $(B - V)\approx 0.32$ 
at $t2$   we obtain  $ E(B - V)\approx 0.33 \pm 0.1$  ~i.e.   
$A_V\approx 1.04\pm 0.3$, assuming $R_V=3.15$. 
Tomov et al. (1999) noted the presence of 
strong NaI D interstellar lines indicative of a large reddening
and consistent with the above value.

At such low  
galactic latitudes ($b=+2^\circ.16$) the use of 
the dust maps of Schlegel et al (1998), to estimate reddening
is 
limited because for $|b|<5^\circ$ the contaminating sources have
not been removed from the maps, so that
they provide at best an upper limit to the the reddening.
The value derived for V 1493 Aql is $E(B-V)=1.67$, which 
appears completely inconsistent with the observed colors.. 
We shall therefore adopt  $ E(B - V)=0.33$ and $A_V =1.04$.

We use three  methods to estimate the absolute
magnitude of the nova in outburst:

\begin{enumerate}

\item
the empirical relation  between
absolute magnitude at maximum and
rate of decline (MMRD, Zwicky 1936; Della Valle \& Livio 1995).
Taking  $\log t2 \sim 0.5$, ( which is likely to be  an upper 
limit) from the MMRD relation of Della Valle \& Livio (1995),
we obtain  $M_V \approx -8.97$.
We point out  that at very short $t2$ , as in our case, the
relation is rather flat  and
produces changes on $M_V$ less than  0.1 mag.
The observed maximum was at  $V= 8.8$ but we allow for a 
slightly brighter $V$ at maximum ( $V \approx 8.3$), since  it 
seems reasonable that the real maximum  has been missed.
Adopting $A_V=1.04$ it is straightforward to obtain $d = 17.6 \pm  
3.2$Kpc.

\item
The absolute
magnitude 15 days  after maximum light appears to be independent 
of the speed class (Buscombe \& de Vaucoleurs 1955)   and is 
a good standard candle. 
As absolute magnitude at day 15 we take $M_V(15) \approx -5.47 
\pm 0.2$ as the  average value of
various estimates reported in Warner (1995, p.266). 
At day 15  we have  $V=12.6 $ implying  $d= 25.5\pm 3.5 $ Kpc.
We point out that day 15 lies inside  the secondary maximum.
Had we used the $V \sim  13.6$ value corresponding with the 
smooth (unperturbed) decline then  $d = 40.4 $ Kpc.

\item
 Another estimate of the nova distance can be 
obtained on the basis of the ``theoretical'' assumption that the
nova
luminosity  at maximum is close to or exceeds the Eddington
luminosity ($L_{edd}$) 
for a 1 ${\rm M_\odot}$ object. In this framework  fast novae 
should reach the  highest peaks, up to $4.7\times10^5 L_\odot$ 
(Warner 1995). If we take $L_{max} \sim 6 L_{edd} \sim  2\times 
10^5 L_\odot$ as a representative 
value for V 1493 Aql, we obtain   
M$_{bol} = -8.55$. 
Near maximum light novae radiate mostly in the
optical and the bolometric  correction BC
is quite small and close to $-0.2$. If we take $M_V = -8.35 $
together with $A_V=1.04$ and
 $V=8.3$ we obtain  $d = 13.2 $Kpc.  

\end{enumerate}
   
It is disturbing  that the absolute visual magnitude 
obtained with the MMRD method ($M_V = -8.95$) is
smaller than the absolute bolometric magnitude 
obtained in the assumption $L \sim  6 L_{edd} (M_{bol} = -8.5)$.
We recall however that the intrinsic dispersion in the MMRD 
curve of Della Valle and Livio (1995) is of the order of $0.4$ 
magnitudes. The two 
estimates can be reconciled  if a bolometric luminosity 
close to 10 $L_{edd}$ is assumed (this gives  $M_V = -8.95$, 
if $BC \sim -0.2$ ). We recall that Duerbeck (1981) indicated 
that very fast novae should radiate at this high rate . Also 
Livio (1992), on the basis of numerical calculations,   
suggested   $L_{max}/L_{edd} \sim 4.63\times M_{wd}^3$. 
Since a rather massive 
white dwarf ($\ge 1.2$ ${\rm M_\odot}$) 
is theoretically required to produce a 
very fast nova such as V 1493 Aql, then also Livio's relation 
points to  $L_{max}/L_{edd} \sim  10$. 
By these arguments the distance  
$d = 13.2 $Kpc obtained with method 3. should be 
considered as a lower  limit and the unweighted average, 
$d =  18.8 \pm 3.6$ Kpc  of the three different estimates is 
rather conservative.

We have tried to estimate the total amount of the ``excess'' 
energy associated with the secondary outburst by measuring the 
area of the region delimited  by the observed luminosity curve 
and the  curve corresponding to a smooth, near exponential 
decline.   The total time interval of the secondary  brightening 
is  40 days and  the extra energy contained in the bump 
associated with  the secondary outburst is of the order of 
$2\times 10^{43}$ erg . We point out, however, that this value 
may be affected by  a non negligible error because of the  
uncertainty in the estimate of the curve corresponding to the 
``smooth'' decline.

\section{Discussion}

The very fast character of the light curve  of V 1493 Aql
associated with the conspicuous re-brightening  
(peak increase in $V \sim 2.5$ mag  with
respect to the smoothed V curve and total duration 
of about 40 days) is not common in novae. 
This, together with the fact that
V 1493 Aql is hotter at maximum light  
($(B-V)<0.4$) than in the following phases: $(B-V)\sim  1$ near 
the secondary maximum  and $\sim  0.7$ in the following 
decline)  makes V 1493 Aql quite peculiar among novae.
We considered the possibility that the rapid reddening of V 1493 Aql
in the early phases
might be due to the formation of dust, however we regard this
unlikely
because dust formation is associated with the presence
of a  ``deep minimum'' in the light curve, which is not observed.
Moreover dust is generally found in slow novae.
 
The large 
distance   we have derived
requires a comment.
Clearly
 $d$  would be  significantly reduced  if the 
apparent magnitude at maximum  were much brighter than 8.8.
If $V_{max}\sim  6.0$ 
the distance would be reduced to  a more
comfortable 6.2 Kpc.   However, we consider  it  very
unlikely  that a $V = 6$ object might have escaped the attention 
of vigilant
sky-watchers during a few days.  The lack of
detection on  films  taken on 9.9 July UT  (Tago 1999),
implies that even if the
nova managed to escape the attention of the astronomers,
it had to decline by $\sim  3$ mag 
in an extremely  short time. Considering that the 
observed decline already suggests a $t2$ less than 
3 days, this would imply a decay by about 5 magnitudes in less
than 6 days, thus making of V 1493 Aql  the fastest known 
nova. 
The  photometric and spectroscopic
behaviour  of V 1493 Aql is
that of a nova; the presence of an emission line spectrum near  
the maximum of the outburst 
(Ayani \& Kawabata 1999; Tomov et al 1999;
Lynch et al 1999)  is generally associated  with the
very fast character in the light curve. 
We are therefore reluctant to 
accept  that this  is  a unique object which does not obey  the 
empirical MMRD-like relations found for other members of its 
class. 

The three  distance estimates are compatible, within 1.7 
$\sigma$. Although the distances estimated
from the magnitude at maximum are uncertain, to the extent
that the ``true'' maximum may have been missed, the fact that
assuming $V_{max}=8.3$ we obtain distances which are compatible
with the estimate from $V$ at day 15, suggests that
the maximum may not have been missed by more than a few tenths of magnitude.

A distance  of $18.8 $ Kpc   
places V 1493 Aql  at  over 
14 Kpc from the Galactic centre, thus the star is at the 
outskirts of our own galaxy.
If however the distance is  25.5 Kpc,
as derived from the magnitude at day 15, then
the  distance from the Galactic centre is 
about 20 Kpc and thus the nova is outside the Galaxy.
In this case it would fall in a 
Local Group galaxy, which could be called the Aquila galaxy.
That a Local Group galaxy at such low Galactic latitudes could
have 
gone undetected is not implausible, so the
fact that no Local Group galaxy is known in that direction
does not allow to rule out this possibility. 
The existence of the Aquila galaxy may be disproved 
by number counts and/or
radial velocity surveys.

\begin{acknowledgements}
We are grateful to G. Carbonaro, A. Di Stefano and M. Puleo
for the assistance during the observations. We wish to thank 
G. Masi for kindly providing an unpublished measure.
We acknowledge with thanks use of the data from the AAVSO International 
Database .
%THANKS TO THE AAVSO and the JAPS

\end{acknowledgements}

\end{document}